\documentstyle[prl,aps,preprint,axodraw]{revtex}
\tightenlines
\input{epsf.sty}
\input{psfig.sty}

\newcommand{\be}{\begin{equation}}
\newcommand{\ee}{\end{equation}}
\newcommand{\bea}{\begin{eqnarray}}
\newcommand{\beas}{\begin{eqnarray*}}
\newcommand{\eea}{\end{eqnarray}}
\newcommand{\eeas}{\end{eqnarray*}} 
\newcommand{\ba}{\begin{array}}
\newcommand{\ea}{\end{array}}
\begin{document}

\draft
\preprint{\vbox{
\hbox{UMD-PP-00-099}}}

\title{Neutrino Mass, Bulk Majoron and Neutrinoless Double Beta Decay}
\author{R. N. Mohapatra$^1$\footnote{e-mail:rmohapat@physics.umd.edu},
A. P\'erez-Lorenzana$^{1,2}$\footnote{e-mail:aplorenz@Glue.umd.edu} 
and C. A. de S. Pires$^{1}$\footnote{e-mail:cpires@physics.umd.edu}}
\address{
$^1$ Department of
Physics, University of Maryland, College Park, MD, 20742, USA\\
$^2$  Departamento de F\'\i sica,
Centro de Investigaci\'on y de Estudios Avanzados del I.P.N.\\
Apdo. Post. 14-740, 07000, M\'exico, D.F., M\'exico. }

\date{August, 2000}

\maketitle
\begin{abstract}
{A new economical model for neutrino masses is proposed in the context of
the brane-bulk
scenarios for particle physics, where the global $B-L$ symmetry of the
standard model is broken spontaneously by a gauge singlet Higgs field in
the bulk. This leads to a bulk singlet majoron whose Kaluza-Klein
excitations may make it visible in neutrinoless double beta
decay for some parameter range if the string scale is close to a TeV.}

\end{abstract}
\pacs{14.60.Pq; 14.60.St.}

\section{Introduction}
One of the major phenomenological challenges for models with large extra
dimensions and low string scale\cite{arkani} is to understand the small
mass of neutrinos. The basic
problem arises due to the fact that the effective theory below the string
scale will have nonrenormalizable operators which are suppressed by powers
of the string scale $M$. The operator relevant for neutrino masses is
of the form $LHLH/M$, where $L$ and $H$ are the lepton
and Higgs doublets of the standard model respectively. After symmetry
breaking, it leads to neutrino masses which are much too large. In
general,
higher dimensional operators can also create other phenomenological
difficulties for such models e.g.
 rapid proton decay via operators of the form $QQQL/M$; however, it has
been suggested that operators that involve different matter fields such as
the ones that lead to proton decay, can be suppressed by using the idea of
``fat'' branes\cite{sch} where different matter fields are located at
different points in the brane. This idea, however,
does not help in the case of the neutrino mass operator
above since it involves only one matter field and since
the Higgs field needs to be ``spread out'' rather than localized for
all fermions to have mass. One must therefore seek other ways to
suppress the effects of this operator.

 A simple way to understand small neutrino masses in these models, 
suggested early on, is to assume the existence of a global $B-L$ symmetry
and include only bulk neutrinos in addition to the standard
model particles\cite{dienes}. This leads to small
neutrino masses for natural values of all parameters, due to suppressed 
overlap of the wave function between the
brane and the bulk fields. The neutrinos in this model are Dirac
particles. A second suggestion is to use a local
 $B-L$ symmetry\cite{mnp}, which generically requires the string scale
to be intermediate rather than TeV type. The neutrinos in this model can
be either  Dirac or  Majorana particles. By now, the phenomenology of
the former case has been studied extensively in several
papers\cite{barbieri}.

More recently an alternative suggestion has been put forth\cite{ma}
where the global $B-L$ symmetry of the standard model is assumed to be an
exact symmetry of the complete model so that all undesirable higher
dimensional terms contributing to neutrino mass are 
forbidden. However, instead of adding extra neutrinos to the bulk,
a scalar
field in a separate brane is used to break the $B-L$ symmetry
spontaneously. This leads to a singlet majoron\cite{cmp}, which
practically decouples from the theory even though the scales are very
small. The smallness of neutrino masses in this model arise from the
Yukawa like suppression (called ``shining''\cite{dimo}) that has its 
origin in the
propagator of a massive bulk field (denoted by $\chi$). In order to
implement this picture, one needs the number of large extra dimensions to
be at least three, preferably more. The reason for at least three extra
dimensions is that the desired suppression takes place only in these
cases. Furthermore, four or more are preferable because
 in case of three extra dimensions, the relation
 \be
 M_{P\ell}^2=M^{5}R^3,
 \label{planckrelation}
 \ee
 implies that the size of the extra dimension is $R\leq
(KeV)^{-1}$ for $M\geq 1$ TeV. Since
to get small neutrino masses via the ``shining'' effect one needs 
$m_{\chi}\ll R^{-1}$, the bulk field must have a tiny mass, much less
than a keV. If the number of
extra dimensions is four or more, this constraint on the parameter $m_{\chi}$
becomes much weakened, requiring less ``fine tuning'' in the theory. The
other point is that even for three extra dimensions, their sizes become so
small that the current gravity experiments lose their usefulness as ways
to search for their existence.

In this paper, we like to pursue the idea that spontaneous breaking of
$B-L$ symmetry may indeed be the origin of neutrino masses in models with
large extra dimensions but with sizes of extra dimensions in the
millimeter
range and furthermore, we will work only with one large extra dimension.
As far as the string scale $M$ goes, we will assume
that $M\sim 30$ TeV so as to satisfy the SN1987A bounds\cite{sn} for the
case of one large extra dimension. Sizes of other dimensions will be
accordingly adjusted so as to satisfy the generalized version of the 
relation in Eq. (1). In order to
get nonvanishing neutrino masses, we will assume that the $B-L$ symmetry
is spontaneously broken by {\it a gauge singlet scalar field $\chi$ in the
bulk}. This scalar field $\chi$ which is assumed to carry two units of 
lepton number ($B-L$), is the only extra
field in the model, thus making it the most economical extension of the
standard model to date that leads to neutrino masses.  (Note in
contrast that in the bulk neutrino  alternative, one
needs a minimum of three bulk fermions to get a realistic mass pattern).
 
An interesting feature of this model is that in some extreme domains of
the parameter space
of the model, the singlet majoron has a chance to be visible in processes
such as neutrinoless double beta decay due to its many Kaluza-Klein
excitations. A
disadvantage is that to get neutrino masses in the eV range, some
suppression of the strengths of the higher dimensional operators or a
small value for $m^2_{\chi}$ is
needed, unlike the bulk neutrino models. The required suppressions are at
the same level as that needed for example in the triplet majoron
model\cite{gel,georgi}.

\section{Bulk singlet and neutrino mass}
Our scenario consists of standard model in the brane, to which
 we add only one gauge singlet complex scalar field $\chi$
propagating in the bulk. We assume that (i) the field $\chi$ carries two
units of $B-L$ quantum number; (ii) the model respects
global $B-L$ symmetry prior to symmetry breaking by vacuum and (iii) it 
includes in the Lagrangian operators of all
dimensions that conserve $B-L$ symmetry. A list of some of the leading
operators are: 
\be
{\cal L}(x)= \int dy\left[
\frac{f}{M^{5/2}}(LH)^2\chi(x,y)+\frac{f'}{M^{17/2}} 
Q^6\tilde{H}^2\chi^*(x,y)\right] \delta(y),
\label{minimaloperator}
\ee
where $\tilde{H}=i\tau_2 H^*$.
We have not included any operator that could be suppressed by appropriate
``fattening'' of the brane\cite{sch}. The second operator in Eq. (2) 
is also not suppressed in the fat brane scenario and is a $\Delta B=2$
operator that can give rise to the process of neutron-anti-neutron
oscillation\cite{marshak}. We will discuss this later.

In order to implement spontaneous breaking of $B-L$ symmetry, let us write
down the bulk scalar potential for $\chi$. The part 
of the potential important here is
\be
V(\chi)=-\frac{m^2_\chi}{2} \chi^{\dagger}\chi +
\frac{\lambda}{4M}(\chi^{\dagger}\chi)^2 .
\label{chipotential}
\ee
Minimizing this potential, we find that at its minimum, the
singlet field has the vev
\be
\langle \chi \rangle_B=\frac{m_\chi M^{1/2}}{\sqrt{\lambda}}.
\label{chiminimumbrane}
\ee
Using this, we find that after electroweak symmetry breaking i.e.
$\langle H \rangle=v_{wk}$, the
neutrinos acquire a Majorana mass given by:
\be
m_\nu=\frac{f}{M^{5/2}} v^2_{wk}\langle \chi \rangle.
\label{neutrmass1}
\ee
For $M=30$ TeV, we can generate neutrino mass of order of 
eV  if we take $\langle \chi \rangle=5\times \left(\frac{10^{-3}}{f}\right)$
GeV$^{3/2}$.
This leads to $m_{\chi}\simeq 90\times \left(\frac{10^{-3}}{f}\right)$ MeV
(for $\lambda \sim 10$).
Thus we need somewhat of a strong fine tuning of the parameters of the
bulk fields  to get the right order for neutrino masses. This fine tuning
is at the same level as that required in the case of the triplet majoron
model\cite{gel}. Despite this feature, we consider these models to be of
interest since
they appear rather economical and embody a new phenomenon not hitherto
discussed in the context of neutrino masses. Furthermore, the small
$m^2_{\chi}$ values could perhaps be made natural if there is
supersymmetry, while keeping the other features unaffected.
As we discuss below, the extreme small mass range of $\chi$ has one
advantage that it makes the associated goldstone boson, the majoron more
visible in certain low energy processes. 

It is clear that due to spontaneous breaking of $B-L$ symmetry in the
bulk, this model has the massless particle, majoron (the CP-odd part of
the singlet $\chi$, denoted by $J$), which is a bulk field. In
four dimensions, the 
majoron has a tower of partners with masses separated by a tiny amount
($\sim 10^{-3}$ eV) for millimeter extra dimensions. They will be
produced as a whole tower in any process where majoron is produced. 
Furthermore, the real part of the field $\chi$(to be denoted by
$\sigma_{\chi}$) also has a mass
$\frac{m^2_{\chi}}{2}$. Since in this model $m_{\chi}$ has a value in
the range of few MeV or less, it and its tower could also be produced in 
processes that have enough phase space. We give
the example of the neutrinoless double beta decay in the next section,
where only the majoron is produced unless the $\sigma_{\chi}$ has a mass
in the sub MeV range. In processes such as muon decay however, both
particles will be produced although the amplitude for it is highly
suppressed.

The neutrino mass texture in this model arises purely from the flavor
profile of the higher dimensional coupling $f_{ij}$. Experimental data on
neutrino oscillations will fix this profile. As an example, which embodies
the so-called bimaximal neutrino mixing pattern and nearly degenerate
neutrino masses, we provide the following $f$ matrix:
\be f~=~\left(\begin{array}{ccc} m_0 & \frac{cs}{\sqrt{2}}\delta_S &
\frac{cs}{\sqrt{2}}\delta_S \\ \frac{cs}{\sqrt{2}}\delta_S &
m_0+\delta_A/2 & \delta_A/2 \\\frac{cs}{\sqrt{2}}\delta_S & \delta_A/2 &
m_0+\delta_A/2 \\ \end{array}\right)
\ee
where $m_0$ is the common mass, $c= cos\theta$ and $s= sin\theta$;
$\delta_{A,S}$ are responsible for the mass splittings that explain the
atmospheric and solar neutrino data. For $c=s=\frac{1}{\sqrt{2}}$, we get
the bimaximal pattern. An advantage of the mass degeneracy is that it
enhances the contribution to the neutrinoless double beta decay.

\section{Neutrinoless double beta decay with majoron emission}
One of the primary experimental manifestation of the majoron idea is in
the process of neutrinoless double beta decay with majoron emission, a
fact which was first noted for the case of the triplet majoron\cite{gel}
in ref.\cite{georgi}. Note that the original singlet majoron coupling
to neutrinos is so weak that it is generally not visible in this process.
However, the bulk majoron, though a gauge singlet, is different. It
can be produced in neutrinoless double beta decay via the diagram in
Fig. 1.

\begin{center}
\begin{picture}(200,50)(0,0)
    \ArrowLine(40,10)(100,10) \ArrowLine(100,10)(160,40)
    \Photon(100,10)(120,-20){3}{8} \ArrowLine(120,-40)(120,-20)
    \ArrowLine(120,-40)(120,-60) \DashLine(120,-40)(170,-40){4}
    \DashLine(120,-40)(90,-40){4} \Photon(120,-60)(100,-90){3}{8}
    \ArrowLine(40,-90)(100,-90) \ArrowLine(100,-90)(160,-120)
    \ArrowLine(120,-20)(180,-20) \ArrowLine(120,-60)(180,-60)
    \Text(70,20)[r]{\large $d_L$} \Text(130,35)[r]{\large $u_L$}
    \Text(70,-100)[r]{\large $d_L$} \Text(130,-115)[r]{\large $u_L$}
    \Text(105,-10)[r]{ $W^-$} \Text(105,-70)[r]{ $W^-$}
    \Text(115,-30)[r]{ $\nu_L$} \Text(115,-50)[r]{ $ \nu_L $}
    \Text(160,-30)[r]{\large $e_L$} \Text(160,-50)[r]{\large $ e_L $}
    \Text(180,-40)[r]{ $ J $} \Text(95,-40)[r]{\large $\times$}
    \Text(85,-40)[r]{ $ \langle H \rangle^2$}
    \Vertex(100,10){1.5} 
    \Vertex(120,-20){1.5}  \Vertex(120,-60){1.5}
    \Vertex(100,-90){1.5}\Vertex(120,-40){1.5}
    \end{picture} 
    \vskip25ex 
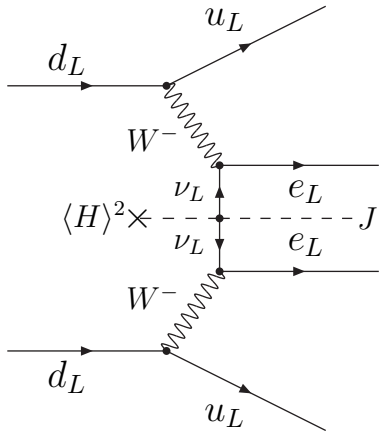
{\sl Fig. 1: Majoron emission in neutrinoless double
 beta decay in the bulk majoron model } \end{center}

The differential decay width for this process can be written as:
\be
\frac{d^2\Gamma}{d\epsilon_1d\epsilon_2}~=~A^2_{Nucl}\frac{G^4_Fp^2_F}{8\pi^5}
\left(\frac{f^2v^4_{wk}}{M^5}\right)(E-\epsilon_1-\epsilon_2)^2\epsilon_1
k_1\epsilon_2 k_2 ,
\ee
where $\epsilon_{1,2}$ and $k_{1,2}$ are the electron energy and momenta
respectively, $A_{Nucl}$ is a dimensionless nuclear factor, whose value we
take from nuclear calculations for the single majoron decay
mode\cite{klap}; $p_F$ is the Fermi
momentum in nuclei (which we take to be 100 MeV); $E$ is the available
energy for electrons and the majoron in the decay\cite{klap}. Note that
there are two
powers of the  factor $(Q-\epsilon_1-\epsilon_2)$ (the power of this
factor in the differential decay distribution is called in the literature
as spectral index\cite{cli}) above in contrast with a
single power in the triplet majoron model and generally odd powers in
most theoretical models\cite{rabi}. This is the effect of the tower of
 majoron KK modes. The lower limits on the lifetime for the process
$\beta\beta_{0\nu J}$ from various nuclei are now at the level of
$7.2\times 10^{20}$ years for $^{48}$Ca\cite{bara} to $7.2\times 10^{21}$
yrs for $^{130}$Xe\cite{lus} and $^{76}$Ge \cite{gun}. For
$A_{Nucl}\sim 0.1$ and $f\sim 1$, using the best of the above experimental
limits\cite{klap}
for majoron emission in
$\beta\beta_{0\nu J}$, we get a lower limit on $M\geq 1$ TeV.
This bound can be improved once the search for the majoron
emitting double beta decay is carried out at higher precision level by
experiments such as for instance the proposed GENIUS\cite{genius}
experiment as well as others.

\section{Neutron-Anti-neutron oscillation}
Let us now turn to the second operator in Eq. (1), which leads to $\Delta
B=2$ transitions. The strength of this operator is given by
\be 
G_{\Delta B=2}\sim \frac{f'\langle \chi \rangle_B v^2_{wk}}{M^{17/2}}.
\ee  
For the values of $M\simeq 30$ TeV and $\langle \chi \rangle_B= 10- 0.01$ GeV$^{3/2}$ 
discussed above, we get $G_{\Delta B=2} \sim 10^{-32} - 10^{-35}$
GeV$^{-5}$. This translates into an oscillation time for $N-\bar{N}$
from $10^{10}$ to $10^{15}$ sec. after allowing for uncertainties in the
hadronic matrix elements. The lower values are in the range accessible to
a proposed experiment\cite{yuri}.

The same operator also leads to a novel process where an infinity tower of 
KK majorons are emitted in the transition $NN\rightarrow \chi$ or in terms
of actual nuclear transmutation $(Z, A)\rightarrow (Z, A-2) + \chi$. The
width for this process is given by:
\be
\Gamma_{\Delta B=2}\sim M^{-17} m^{14}_N v^4_{wk} 10^{-6} GeV,
\ee
where we have used the factor of $10^{-6}$ to denote the hadron
dressing of the six quark operator. For $M=30$ TeV, the nuclear
instability life times implied by this are $\sim 10^{39}$ yrs. However,
if we ignored other constraints on $M$ and chose it to be of order $10$
TeV or so, we would expect majoron emitting modes of the above type with a
life time of about $10^{32}$ years which looked for perhaps even in
existing data.

\section{conclusion}
In this brief note, we have suggested a new model for neutrino
masses in theories with large extra dimensions using spontaneous breaking
of lepton number symmetry by a bulk scalar field. The resulting bulk
majoron 
may be visible in neutrinoless double beta decay experiments for certain
domains of parameters, if the string scale is indeed close to a TeV. The
model also predicts a novel baryon number violating process
where two neutrons in a nucleus disappear with the emission of a
majoron which would lead to missing energy proton decay events.


{\it Acknowledgements.}   The work of RNM is supported by a grant from the
National Science Foundation under grant number PHY-9802551. The work of
APL is supported in part by CONACyT (M\'exico). The work of CP is
supported by Funda\c c\~ao de Amparo \`a Pesquisa do Estado de S\~ao Paulo
(FAPESP).


\end{document}